# Generalized Spin and Pseudo-Spin Symmetry: Relativistic Extension of Supersymmetric Quantum Mechanics

A. D. Alhaidari

*Saudi Center for Theoretical Physics, Jeddah, Saudi Arabia*

We consider the Dirac equation in 1+1 space-time dimension with vector, scalar and pseudo-scalar coupling. In the traditional spin (or pseudo-spin) symmetry, the difference between (or sum of) the scalar and vector potentials is a constant. Here, however, we introduce an alternative symmetry where the scalar or pseudo-scalar potential is proportional to the vector potential. This leads to a model with significant extensions to supersymmetric quantum mechanics. We present a formal solution of the problem but give explicit analytic results for specific examples.

## I. INTRODUCTION

In nuclear physics, certain aspects of deformed and exotic nuclei have been studied by means of spin symmetric and pseudo-spin symmetric concepts [1,2]. Spin symmetry occurs in the spectrum of mesons with one heavy quark and it is used to explain the absence of quark spin-orbit splitting (spin doublets) [3], which is observed in heavy-light quark mesons. Pseudo-spin symmetry, on the other hand, is used to explain the observed degeneracies of some shell-model orbitals in nuclei (pseudospin doublets) [4], exotic nuclei [5] and in the case of triaxiality as well [6]. Concepts in this symmetry have been used to explain features not just of deformation but also of super-deformation in nuclei [7] and to establish an effective nuclear shell-model scheme [1,2,8]. In real nuclei, however, pseudo-spin symmetry is only an approximation whose quality depends on the associated pseudo-spin orbital potential [9]. Ginocchio [4,10,11] showed that spin symmetry occurs when the difference between the vector potential and scalar potential in the Dirac Hamiltonian is a constant, whereas, pseudo-spin symmetry occurs when the sum of the two is a constant. In the exact spin symmetric limit, the Dirac equation corresponds to a Schrödinger equation with SU(3) symmetry [10]. Pseudo-spin has also been used in the nonrelativistic harmonic oscillator, where it is found that the relation between the coefficients of spin-orbit and orbit-orbit terms in the nonrelativistic single-particle Hamiltonian is consistent with those obtained in the relativistic mean-field theory [11,12].

In all past developments on spin and pseudo-spin symmetry, aside from the work of Leviatan [13], it is rare, if ever, that full supersymmetry is demonstrated and utilized. And when so, it is only limited to the use of techniques of supersymmetric quantum mechanics in the solution of shape invariant potentials [14]. In the conventional spin and pseudo-spin symmetry, the scalar and vector potentials of the problem are related by $S = \pm V + q$, where $q$ is a constant. The resulting dynamics of such a system as exhibited by the structure of the associated differential equation [see Eq. (8) in the text below] does not exemplify the well-known features of the supersymmetric theory. In this work, however, we generalize the symmetry and investigate its consequences in simple 1+1 space-time dimensional models. Generalization is achieved by proposing the following alternative relation between the potentials: $S = \lambda V + q$, where $\lambda$ is a dimensionless parameter. We show that this generalization endow the theory with supersymmetry, but with significant extensions over the traditional theory [see Eq. (11a) in the text and the ensuing discussion that follows]. Additionally, we consider the case where a pseudo-scalar coupling is introduced such that $W = \lambda V$, where $W$ is the pseudo-scalar potential.

In the following section, we formulate the problem by constructing the Dirac Hamiltonian in 1+1 space-time with the most general potential coupling that includes vector, scalar, and pseudo-scalar potentials. We demonstrate that the proposed generalization of the traditional (spin and pseudo-spin) symmetry will entail non-trivial relativistic extension of supersymmetric quantum mechanics. In Sec. III, we illustrate these remarkable qualities by showing the detailed analysis and giving explicit results for the Morse potential model. In Sec. IV, we provide results without details for two more interesting examples. We conclude by giving an outline of the problem in 3+1 dimensions.

## II. FORMULATION

In the conventional relativistic units, $\hbar = c = 1$, the Dirac equation for a free structureless fermion of mass $m$ reads as follows:

$$\left(\gamma^\mu p_\mu - m\right)\Psi = \left(i\gamma^\mu \partial_\mu - m\right)\Psi = 0, \tag{1}$$

where repeated indices are summed over. In 1+1 space-time, $\{\gamma^\mu\}_{\mu=0}^{1}$ are two constant square matrices satisfying the anti-commutation relation $\{\gamma^\mu, \gamma^\nu\} = 2\eta^{\mu\nu}$. The Mikowski space-time metric is $\eta = \text{diag}(+-)$ and the 2-gradiant is $\partial_\mu = \left(\frac{\partial}{\partial t}, \frac{\partial}{\partial x}\right)$. Gauge invariant coupling of the fermion to the 2-vector potential $A_\mu$ is accomplished by the minimal substitution $p_\mu \to p_\mu - gA_\mu$, where $g$ is the coupling parameter. We write the 2-vector potential $A_\mu = (V, U)$, in terms of which we can rewrite the Dirac equation (1) as

$$\left(i\gamma^\mu \partial_\mu - g\gamma^\mu A_\mu - m\right)\Psi = 0, \tag{2}$$

which is invariant under the gauge transformation: $A_\mu \to A_\mu + \partial_\mu \Lambda$, $\Psi \to e^{-ig\Lambda}\Psi$, and $\Lambda(t,x)$ is an arbitrary scalar function. For time independent potentials, we can write the multi-component wavefunction as $\Psi(t,x) = e^{-i\varepsilon t}\psi(x)$, where $\varepsilon$ is the relativistic energy. Thus, Eq. (2) becomes

$$\left(\varepsilon + i\alpha \frac{d}{dx} - gV - g\alpha U - m\beta\right)\psi = 0, \tag{3}$$

where $\alpha = \gamma^0 \gamma^1$ and $\beta = \gamma^0$. In even dimensions (e.g., 1+1 and 3+1), all matrix representations of the Dirac $\gamma$ matrices are equivalent [15]. We choose the following standard representation that satisfies the Clifford algebra $\{\gamma^\mu, \gamma^\nu\} = 2\eta^{\mu\nu}$

$$\gamma^0 = \sigma_3 = \begin{pmatrix} 1 & 0 \\ 0 & -1 \end{pmatrix}, \quad \gamma^1 = i\sigma_1 = \begin{pmatrix} 0 & i \\ i & 0 \end{pmatrix}. \tag{4}$$

Then, we can write Eq. (3) as a matrix differential equation for the two components, $\psi^\pm(x)$, of the wavefunction $\psi$

$$\begin{pmatrix} m+gV-\varepsilon & \frac{d}{dx}+igU \\ -\frac{d}{dx}-igU & -m+gV-\varepsilon \end{pmatrix}\begin{pmatrix} \psi^+ \\ \psi^- \end{pmatrix}=0 \qquad (5)$$

We can also add a scalar potential $S(x)$ by the substitution $m \to m+S$. Additionally, including a pseudo scalar potential, $W(x)$, we obtain the following Dirac equation with the most general potential interaction in 1+1 dimension

$$\begin{pmatrix} m+S+gV-\varepsilon & \frac{d}{dx}+igU+W \\ -\frac{d}{dx}-igU+W & -m-S+gV-\varepsilon \end{pmatrix}\begin{pmatrix} \psi^+ \\ \psi^- \end{pmatrix}=0. \qquad (6)$$

Taking $W = U = 0$ and imposing the condition that $S = \pm gV + q$, where $q$ is a real constant parameter, result in the usual spin and pseudo-spin symmetry of the problem (depending on the sign). This choice, which is typical of most published work on this symmetry, turns Eq. (6) into

$$\begin{pmatrix} M+(1\pm 1)gV-\varepsilon & \frac{d}{dx} \\ -\frac{d}{dx} & -M+(1\mp 1)gV-\varepsilon \end{pmatrix}\begin{pmatrix} \psi^+ \\ \psi^- \end{pmatrix}=0, \qquad (7)$$

where $M = m+q$. This relates one component of the wavefunction to the other by $\psi^{\mp} = \left(\frac{-1}{m\pm\varepsilon}\right)d\psi^{\pm}/dx$ and gives the following second order differential equation

$$\left[-\frac{d^2}{dx^2}+2g(\varepsilon\pm M)V(x)+M^2-\varepsilon^2\right]\psi^{\pm}(x)=0. \qquad (8)$$

Here, we also take $W = U = 0$ but generalize the relation between the potentials as $S = \lambda V + q$, where $\lambda$ is a real dimensionless parameter. For a general $\lambda$, the formal solution of the problem turns out to be complicated. In fact, aside from the constant potential model, the resulting equation is not Schrödinger-like and its solution appears to be intractable. However, it is possible that simplification could result from the action of a special rotation in function space. Consequently, we explore this possibility by applying the global unitary transformation $\exp\left(\frac{i}{2}\theta\sigma_2\right)$ to the Dirac equation (6), where $\sigma_2 = \begin{pmatrix} 0 & -i \\ i & 0 \end{pmatrix}$ and $\theta$ is a real angular parameter. The transformed matrix equation becomes

$$\begin{pmatrix} \mathcal{C}M+(g+\lambda\mathcal{C})V-\varepsilon & \frac{d}{dx}-\mathcal{S}M-\lambda\mathcal{S}V \\ -\frac{d}{dx}-\mathcal{S}M-\lambda\mathcal{S}V & -\mathcal{C}M+(g-\lambda\mathcal{C})V-\varepsilon \end{pmatrix}\begin{pmatrix} \phi^+ \\ \phi^- \end{pmatrix}=0, \qquad (9)$$

where $\mathcal{S}=\sin\theta$, $\mathcal{C}=\cos\theta$ and $\begin{pmatrix}\phi^+ \\ \phi^-\end{pmatrix}=\begin{pmatrix}\cos\frac{\theta}{2} & \sin\frac{\theta}{2} \\ -\sin\frac{\theta}{2} & \cos\frac{\theta}{2}\end{pmatrix}\begin{pmatrix}\psi^+ \\ \psi^-\end{pmatrix}$. The requirement that $\cos\theta = \pm g/\lambda$ with $0 \leq \theta \leq \pi$, takes Eq. (9) into

$$\begin{pmatrix} \pm M\frac{g}{\lambda}+(1\pm 1)gV-\varepsilon & \frac{d}{dx}-\mathcal{S}M-\lambda\mathcal{S}V \\ -\frac{d}{dx}-\mathcal{S}M-\lambda\mathcal{S}V & \mp M\frac{g}{\lambda}+(1\mp 1)gV-\varepsilon \end{pmatrix}\begin{pmatrix} \phi^+ \\ \phi^- \end{pmatrix}=0, \qquad (10)$$

where $\mathcal{S}=\sqrt{1-(g/\lambda)^2}$, which requires that $|\lambda|\geq g$. Equation (10) is equivalent to the following differential equations for the two components of the wavefunction

$$\left[-\frac{d^2}{dx^2}+\lambda^2\mathcal{S}^2V^2\mp\lambda\mathcal{S}\frac{dV}{dx}+2(g\varepsilon+\lambda M)V+M^2-\varepsilon^2\right]\phi^{\pm}=0, \qquad (11a)$$

$$\phi^{\mp} = \frac{-1}{\varepsilon + (g/\lambda)M}\left(\pm\frac{d}{dx} + \lambda \mathcal{S}V + \mathcal{S}M\right)\phi^{\pm}, \tag{11b}$$

where $\varepsilon \neq -\frac{g}{\lambda}M$, which makes these equations valid only in the positive/negative energy subspace if $\lambda$ is positive/negative. Equation (8) is a special case of (11a) with $\lambda = \pm g$ and, hence, $\mathcal{S} = 0$. On the other hand, unlike Eq. (8), the structure of this Schrödinger-like equation has some of the main features associated with supersymmetric quantum mechanics (SSQM) [16]. However, there are two main nontrivial extensions to the non-relativistic theory:

(1) The two potential partners in SSQM are $\mathcal{V}^2 \pm \mathcal{V}'$, where $\mathcal{V}$ is the superpotential, whereas in (11a) there is an additional term proportional to $\mathcal{V}$. Here, $\mathcal{V} = \lambda \mathcal{S}V$.

(2) The overall effective potential in the theory is energy dependent due to the contribution of the term $\varepsilon \mathcal{V}$, which goes to $m\mathcal{V}$ in the nonrelativistic limit.

The similarity to the SSQM problem will be even more evident when we compute the energy spectrum below. The relativistic extension of SSQM exhibited in some models that are governed by the Dirac theory is due to the fact that the associated Dirac Hamiltonian is an element of the 4-dimensional superalgebra U(1/1), which is a $Z_2$ graded extension of SO(2,1) Lie algebra [17].

In this work, however, we consider the case where $S = U = 0$ and require that $W(x) = \lambda V(x)$, where $\lambda$ is a real parameter that might be constrained by the desired symmetry requirements[†]. For the reasons stated above and the supporting results below, we refer to this symmetry as superspin and pseudo-superspin symmetry (depending on the sign of $\lambda$). Again, we apply the transformation $e^{\frac{i}{2}\theta\sigma_2}$ to Eq. (6) but now require that $\sin\theta = \pm g/\lambda$ with $-\frac{\pi}{2} \leq \theta \leq +\frac{\pi}{2}$. The result is as follows

$$\begin{pmatrix} \mathcal{C}m + (1\pm1)gV - \varepsilon & \frac{d}{dx} \mp m\frac{g}{\lambda} + \lambda\mathcal{C}V \\ -\frac{d}{dx} \mp m\frac{g}{\lambda} + \lambda\mathcal{C}V & -\mathcal{C}m + (1\mp1)gV - \varepsilon \end{pmatrix}\begin{pmatrix} \phi^+ \\ \phi^- \end{pmatrix} = 0. \tag{12}$$

where $\mathcal{C} = \sqrt{1 - (g/\lambda)^2}$, which is real and non-negative requiring that $|\lambda| \geq g$. Equation (12) gives the following Schrödinger-like differential equation and differential relation to be satisfied by the two components of the wavefunction

$$\left[-\frac{d^2}{dx^2} + \lambda^2\mathcal{C}^2V^2 \mp \lambda\mathcal{C}\frac{dV}{dx} + 2g\varepsilon V + m^2 - \varepsilon^2\right]\phi^{\pm} = 0, \tag{13a}$$

$$\phi^{\mp} = \frac{-1}{\mathcal{C}m \pm \varepsilon}\left(\frac{d}{dx} + m\frac{g}{\lambda} \mp \lambda\mathcal{C}V\right)\phi^{\pm}, \tag{13b}$$

where $\varepsilon \neq \mp \mathcal{C}m$, which is automatically satisfied for positive/negative energies since $\mathcal{C}$ is positive. The structure of Eq. (13a) is the same as that of Eq. (11a). Thus, it has also some of the features associated with SSQM. However, like Eq. (11a), it embodies significant extensions to the nonrelativistic theory.

Now, to obtain the solution space of the problem with the new symmetry, we solve Eq. (13a) for the top/bottom sign and substituting that in Eq. (13b) with the corresponding sign giving the positive/negative energy solution associated with superspin or pseudo-superspin symmetry (depending on the sign of $\lambda$). Next, we give explicit analytic solutions for three models corresponding to the Morse, Rosen-Morse, and Hulthén potentials. We show the detailed analysis for only one of them, say the Morse [18], and just state the results for the other two.

---

[†] For massless theory, a more appropriate relation is $W(x) = \lambda V(x) + q$, where $q$ is a real parameter of mass dimension.

### III. THE MORSE MODEL

We take $V(x) = V_0 e^{-\mu x}$, where $\mu > 0$ and $x$ is the whole real line. Consequently, Eq. (13a) becomes

$$\left[ -\frac{d^2}{dx^2} + \mu^2 \omega^2 \left( e^{-2\mu x} - 2\rho_\pm e^{-\mu x} \right) + m^2 - \varepsilon^2 \right] \phi^\pm = 0, \tag{14}$$

where $\rho_\pm = -\left( \frac{g\varepsilon}{\lambda \mu C} \pm \frac{1}{2} \right) / \omega$ and $\omega = \lambda C V_0 / \mu$. This equation has an exact analytic solution for the bound states ($|\varepsilon| < m$). Its derivation goes as follows. Let $y = e^{-\mu x} \geq 0$, then Eq. (14) becomes

$$\left[ y^2 \frac{d^2}{dy^2} + y \frac{d}{dy} - \omega^2 y (y - 2\rho_\pm) + \frac{\varepsilon^2 - m^2}{\mu^2} \right] \phi^\pm = 0. \tag{15}$$

We consider the ansatz: $\phi_n^\pm(y) = A_n^\pm y^\tau e^{-y/2} L_n^\nu(y)$, where $L_n^\nu(y)$ is the associated Laguerre polynomial, $A_n^\pm$ is a normalization constant, $\tau > 0$, $\nu > -1$, and $n = 0, 1, 2, \ldots$ This function could support bound states since it is square integrable and satisfies the boundary conditions. Substituting this in Eq. (15) and using the differential equation of the Laguerre polynomials [19], we obtain

$$\left[ (2\tau - \nu) \frac{d}{dy} + \left( \frac{1}{4} - \omega^2 \right) y + \frac{1}{y} \left( \tau^2 + \frac{\varepsilon^2 - m^2}{\mu^2} \right) - \left( n + \tau + \frac{1}{2} - 2\rho_\pm \omega^2 \right) \right] L_n^\nu = 0. \tag{16}$$

Now, since all terms inside the square bracket are linearly independent then we must impose the following conditions

$$\omega^2 = \frac{1}{4}, \quad \nu = 2\tau = \rho_\pm - 2n - 1, \quad \varepsilon^2 = m^2 - \tau^2 \mu^2, \tag{17}$$

giving $\lambda^2 = g^2 + \left( \mu / 2V_0 \right)^2$. The last relation above requires that $\tau \leq m/\mu$, which together with $\tau > 0$ give the bounds on the spectrum. Additionally, the last relation in (17) results in the following equation for the relativistic energy spectrum

$$\left( \frac{\varepsilon}{mC} \right)^2 + \left( \frac{\varepsilon}{mC} \right)\left( \frac{\mu g}{m\lambda} \right)\left( \frac{2n+1}{2\omega} \pm 1 \right) + \left( \frac{\mu}{2m} \right)^2 \left( \frac{2n+1}{2\omega} \pm 1 \right)^2 = 1, \tag{18}$$

which is a quadratic equation in the energy whose solution gives the following energy spectrum formula that depends on the physical parameters $m$, $g$, $\mu$, and $V_0$

$$\varepsilon_n^\pm / mC = -2g \left( CV_0/m \right)\left( n + \frac{1}{2} \pm \omega \right) + \text{sgn}\sqrt{1 - \left( C\mu/m \right)^2 \left( n + \frac{1}{2} \pm \omega \right)^2}, \tag{19}$$

where "sgn" is a $+$ or $-$ sign, which is chosen to keep the formula within the permissible bounds of the spectrum dictated by the condition $0 < \tau \leq m/\mu$. It turns out that this sign should be taken opposite to that of $V_0$. Therefore, the overall sign of the energy spectrum is the negative of that of $V_0$. One can show that the spectrum is bound by $m > \left| \varepsilon_n^\pm \right| \geq mC$ with $\left| \varepsilon_0^\pm \right| = mC$. Since $|\omega| = \frac{1}{2}$, then the $n$-dependent factor in (19) is either $n$ or $n+1$.

Thus, aside from a single eigenstate, whose eigen-energy is $\pm m\mathcal{C}$ and belongs only to one of them, the two types of symmetry are isospectral (i.e., they have the same energy spectrum). This property is typical of super-partner potentials in SSQM [16] and it is one of our justifications, in addition to those noted above, for the name given to this symmetry. The size of the energy spectrum could be obtained from the reality condition of formula (19). That is, $n = 0,1,2,...,n_{max}^{\pm}$ where $n_{max}^{\pm}$ is the largest integer $\leq \frac{m}{\mu\mathcal{C}} - \frac{1}{2} \mp \omega$. Superspin symmetry is the one with the larger size spectrum (larger by one). Since the sign of $\omega$ is the same as that of $\lambda V_0$, then it is evident that $\varepsilon_n^+ = \varepsilon_{n+1}^-$ if $\lambda V_0 > 0$ and $\varepsilon_n^- = \varepsilon_{n+1}^+$ if $\lambda V_0 < 0$. Thus, superspin symmetry corresponds to the top (bottom) signs above if the sign of $\lambda$ is opposite to (same as) that of $V_0$, respectively. An alternative and equivalent, though elegant, energy spectrum formula is obtained in the Appendix.

To find the other component of the eigen-spinor, we substitute the ansatz for the component $\phi_n^{\pm}$ in Eq. (13b) and use the differential property and recursion relations of the Laguerre polynomials [19]. Thus finally, the two components of the wavefunction are

$$\phi_n^{\pm}(y) = A_n^{\pm} y^{\tau_n^{\pm}} e^{-y/2} L_n^{2\tau_n^{\pm}}(y), \tag{20a}$$

$$\phi_n^{\mp}(y) = \frac{\mu A_n^{\pm}}{\mathcal{C}m \pm \varepsilon_n^{\pm}} y^{\tau_n^{\pm}} e^{-y/2} \left[ -\left(\tau_n^{\pm} + \frac{mg}{\lambda\mu}\right) L_n^{2\tau_n^{\pm}}(y) \right. \\ \left. + \delta_{\pm}(n+1) L_{n+1}^{2\tau_n^{\pm}-1}(y) + (1-\delta_{\pm})(n+2\tau_n^{\pm}) L_n^{2\tau_n^{\pm}-1}(y) \right] \tag{20b}$$

where $\tau_n^{\pm} = -\left(2g\,\varepsilon_n^{\pm} \frac{V_0}{\mu^2} \pm \omega\right) - n - \frac{1}{2}$ and $\delta_{\pm} = \frac{1}{2} \mp \omega$, which is either zero or one depending on the sign of $\lambda V_0$. Note that for negative $\lambda$, if $V_0 > 0$ then $\varepsilon_0^+ = -\mathcal{C}m$ and $\phi_0 = \begin{pmatrix} \phi_0^+ \\ 0 \end{pmatrix}$ but if $V_0 < 0$ then $\varepsilon_0^- = +\mathcal{C}m$ and $\phi_0 = \begin{pmatrix} 0 \\ \phi_0^- \end{pmatrix}$. In the following section, we give results without details for two more models: the Rosen-Morse and Hulthén potentials.

## IV. THE ROSEN-MORSE AND HULTHÉN MODELS

For the Rosen-Morse potential model [20], we take $V(x) = V_0 \tanh \mu x$, where $\mu > 0$ and $x$ is the whole real line. The associated potential partners in the Schrödinger-like equation (13a) become

$$V_{eff}^{\pm}(x) = \frac{\mu^2}{2} \left[ \frac{\omega(\pm 1 - \omega)}{\cosh^2 \mu x} + 2g\varepsilon \frac{V_0}{\mu^2} \tanh \mu x \right] + \frac{\mu^2}{2} \omega^2. \tag{21}$$

Repeating the same type of calculation as in the previous section, we obtain the following energy spectrum formula

$$\left(\varepsilon_n^{\pm}/\mu\right)^2 = \left[(m/\mu)^2 + \omega^2 - (\xi_n^{\pm})^2\right] / \left[1 + (gV_0/\mu\xi_n^{\pm})^2\right], \tag{22}$$

where $\xi_n^{\pm} = \left|\omega \pm \frac{1}{2}\right| - n - \frac{1}{2}$. The corresponding component of the wavefunction reads as follows

$$\phi_n^{\pm}(y) = A_n^{\pm} (1-y)^{\frac{1}{2}\alpha_n^{\pm}} (1+y)^{\frac{1}{2}\beta_n^{\pm}} P_n^{(\alpha_n^{\pm}, \beta_n^{\pm})}(y), \tag{23}$$

−6−

where $\alpha_n^\pm = \frac{1}{2}\sqrt{\omega^2 + \mu^{-2}[m^2 - (\varepsilon_n^\pm)^2 + 2gV_0\varepsilon_n^\pm]}$, $\beta_n^\pm = \frac{1}{2}\sqrt{\omega^2 + \mu^{-2}[m^2 - (\varepsilon_n^\pm)^2 - 2gV_0\varepsilon_n^\pm]}$, $y = \tanh\mu x$ and $P_n^{(\alpha,\beta)}(y)$ is the Jacobi polynomial. The spectrum is finite with $n = 0,1,2,...,n_{max}^\pm$, where $n_{max}^\pm$ is the largest integer $\le |\omega \pm \frac{1}{2}| - \frac{1}{2}$. To find the other component of the wave function, we substitute (23) in Eq. (13b) and use the differential property and recursion relations of the Jacobi polynomials [19].

The Hulthén potential model [21] corresponds to $V(x) = V_0/(e^{\mu x} - 1)$, where again $\mu > 0$ but now $x$ is half of the real line ($x \ge 0$). The corresponding effective potential partners are

$$V_{eff}^\pm(x) = \frac{\mu^2}{2}\left[\omega\frac{\omega \pm e^{\mu x}}{(e^{\mu x} - 1)^2} + 2g\varepsilon\frac{V_0/\mu^2}{e^{\mu x} - 1}\right]. \tag{24}$$

The energy spectrum formula associated with this model is calculated as

$$\varepsilon_n^\pm = \frac{\mu}{2}\left[\rho^2 + (\xi_n^\pm)^2\right]^{-1}\left\{\rho\left[\omega^2 - (\xi_n^\pm)^2\right] + \text{sgn}\,\xi_n^\pm\sqrt{\left(\frac{2m}{\mu}\right)^2[\rho^2 + (\xi_n^\pm)^2] - [\omega^2 - (\xi_n^\pm)^2]^2}\right\}, \tag{25}$$

where $\rho = gV_0/\mu$, $\xi_n^\pm = n + \beta^\pm$, and

$$\beta^\pm = \begin{cases} \omega + \frac{1\pm 1}{2} & , \lambda V_0 > 0 \\ -\omega + \frac{1\mp 1}{2} & , \lambda V_0 < 0 \end{cases} \tag{26}$$

The "sgn" in (25) is the negative of the sign of $V_0$ (i.e., $\text{sgn} = -V_0/|V_0|$). Therefore, here too as in the Morse model, the sign of the energy spectrum is the negative of that of $V_0$. The corresponding energy eigen-function, $\phi_n^\pm$, is obtained as follows

$$\phi_n^\pm(y) = A_n^\pm y^{\alpha_n^\pm}(1-y)^{\beta^\pm}P_n^{(2\alpha_n^\pm, 2\beta^\pm - 1)}(1 - 2y), \tag{27}$$

where $y = e^{-\mu x}$ and $\alpha_n^\pm = \mu^{-1}\sqrt{m^2 - (\varepsilon_n^\pm)^2}$. The index $n$ runs from 0 to $n_{max}^\pm$, where $n_{max}^\pm$ is the largest integer $\le -\beta^\pm + |\omega|\left\{1 + 2(m/\mu\omega)^2\left[1 + \sqrt{1 + (\lambda V_0/m)^2}\right]\right\}^{1/2}$. The other component of the eigen-function, $\phi_n^\mp$, is calculated by substituting (27) in Eq. (13b) and using the differential property and recursion relations of the Jacobi polynomials.

## V. CONCLUSION AND DISCUSSION

Spin and pseudo-spin symmetry found interesting applications in nuclear physics and other fields. In such applications, the scalar and vector potentials of the relativistic problem are related in a simple way. We have shown that a generalization of this symmetry as outlined above resulted in very fruitful outcome. The new theory not only acquired additional supersymmetry but gained a non-trivial relativistic extension of the features associated with that symmetry as well. In the present work, we illustrated some of these advantages in simple relativistic models in 1+1 dimension. We found that the structure of the solutions of these models (energy spectra and function space) was greatly enriched. Further development should address the problem in 3+1 dimensions, where the dynamics is described by the following equations that replace Eqs. (11)

$$\left[-\vec{\nabla}^2 + \lambda^2 S^2 V^2 \mp \lambda S\vec{\sigma}\cdot(\vec{\nabla}V) + 2(g\varepsilon + \lambda M)V + M^2 - \varepsilon^2\right]\chi^\pm = 0, \tag{28a}$$

$$\chi^{\mp} = \frac{-1}{\varepsilon + (g/\lambda)M}\left(\pm\vec{\sigma}\cdot\vec{\nabla} + \lambda\mathcal{S}V + \mathcal{S}M\right)\chi^{\pm}, \qquad (28b)$$

where $\chi^{\pm}(\vec{r})$ is now a two-component spinor and $\psi(\vec{r}) = e^{-\frac{i}{2}\theta\sigma_2}\begin{pmatrix} i\chi^+ \\ \chi^- \end{pmatrix}$.

**ACKNOWLEDGMENT:** The author is grateful to the Saudi Center for Theoretical Physics (SCTP, Dhahran and Jeddah) for support.

# APPENDIX
# ALTERNATIVE ENERGY SPECTRUM FORMULA FOR THE MORSE MODEL

The last relation in (17) could be written as follows

$$\left(\frac{\varepsilon}{m}\right)^2 + \left(\frac{\mu}{2m}\right)^2\left(2n+1\pm 2\omega + \varepsilon\frac{4g\omega}{\lambda\mu\mathcal{C}}\right)^2 = 1. \qquad (A1)$$

For bound states $-m \leq \varepsilon \leq +m$, thus we can always define an angle $\varphi$ such that $\varepsilon/m = \cos\varphi$. Consequently, we could rewrite Eq. (A1) as

$$\left(\frac{\mu}{2m}\right)^2\left(2n+1\pm 2\omega + \cos\varphi\frac{4gmV_0}{\mu^2}\right)^2 = \sin^2\varphi. \qquad (A2)$$

For all values of the physical parameters, we can also define another angle[‡] $\kappa$ such that $\tan\kappa = 2gV_0/\mu$. In terms of these two angles, the square root of Eq. (A2) reads as follows

$$\left(\frac{\mu}{m}\right)\left(n+\tfrac{1}{2}\pm\omega\right) + \cos\varphi\tan\kappa = \delta\sin\varphi, \qquad (A3)$$

where $\delta$ is a + or − sign. Therefore, we can write this as

$$\sin\kappa\cos\varphi - \delta\cos\kappa\sin\varphi = -\left(\frac{\mu}{m}\right)\left(n+\tfrac{1}{2}\pm\omega\right)\cos\kappa. \qquad (A4)$$

Or, equivalently, $\sin(\kappa - \delta\varphi) = -\left(\frac{\mu}{m}\right)\left(n+\tfrac{1}{2}\pm\omega\right)\cos\kappa$, which gives the following elegant energy spectrum formula

$$\varepsilon_n^{\pm} = m\cos\left\{\kappa + \sin^{-1}\left[\left(\frac{\mu}{m}\right)\left(n+\tfrac{1}{2}\pm\omega\right)\cos\kappa\right]\right\}. \qquad (A5)$$

One can show that this is equivalent to the formula (19). However, this one gives a more transparent picture of the overall property of the spectrum, its bounds, and general behavior with the bound state index $n$.

---

[‡] This definition determines the angle $\kappa$ modulo $\pi$. It turns out that for positive values of $V_0$, one needs to add a $\pi$ to $\kappa$.